\documentclass{nature}
\usepackage[pdftex]{graphicx}
\usepackage{float}
\usepackage{url}

\usepackage{times}
\usepackage{ulem}
\usepackage{threeparttable}

\usepackage{lineno}
\title{Probable detection of an eruptive filament from a superflare on a solar-type star}

\author{Kosuke Namekata$^{1,2,3\ast}$, 
Hiroyuki Maehara$^{4}$, 
Satoshi Honda$^5$,
Yuta Notsu$^{6,7,8}$,
Soshi Okamoto$^{1}$, 
Jun Takahashi$^{5}$, 
Masaki Takayama$^{5}$,
Tomohito Ohshima$^{5}$,
Tomoki Saito$^{5}$, 
Noriyuki Katoh$^{5,9}$,
Miyako Tozuka$^{5}$, 
Katsuhiro L. Murata$^{10}$, 
Futa Ogawa$^{10}$, 
Masafumi Niwano$^{10}$,
Ryo Adachi$^{10}$, 
Motoki Oeda$^{10}$,
Kazuki Shiraishi$^{10}$,
Keisuke Isogai$^{2,11}$, 
Daikichi, Seki$^{1,2,12}$,
Takako T. Ishii$^{2}$,
Kiyoshi Ichimoto$^{2}$, 
Daisaku Nogami$^{1}$ 
\& Kazunari Shibata$^{13, 14}$}

\begin{document}

\maketitle
\begin{affiliations} 
 \item Department of Astronomy, Kyoto University, Sakyo, Kyoto, Japan
 \item Astronomical Observatory, Kyoto University, Sakyo, Kyoto, Japan 
 \item ALMA Project, NAOJ, NINS, Osawa, Mitaka, Tokyo, Japan
 \item Okayama Branch Office, Subaru Telescope, National Astronomical Observatory of Japan, NINS, Asakuchi, Okayama, Japan
 \item Nishi-Harima Astronomical Observatory, Center for Astronomy, University of Hyogo, Sayo, Hyogo, Japan
 \item Laboratory for Atmospheric and Space Physics, University of Colorado Boulder, Boulder, CO, USA
 \item National Solar Observatory, Boulder, CO, USA
 \item Department of Earth and Planetary Sciences, Tokyo Institute of Technology, Meguro, Tokyo, Japan
 \item Graduate School of Human Development and Environment, Kobe University, Nada, Kobe, Japan
 \item Department of Physics, Tokyo Institute of Technology, Meguro, Tokyo, Japan
 \item Department of Multi-Disciplinary Sciences, Graduate School of Arts and Sciences, The University of Tokyo, Meguro, Tokyo, Japan
 \item Graduate School of Advanced Integrated Studies in Human Survivability, Kyoto University, Sakyo, Kyoto, Japan
 \item Kwasan Observatory, Kyoto University, Yamashina, Kyoto, Japan
 \item School of Science and Engineering, Doshisha University, Kyotanabe, Kyoto, Japan
\end{affiliations}

%\normalsize{$^\ast$To whom correspondence should be addressed; E-mail:  namekata@kusastro.kyoto-u.ac.jp.}

%\clearpage

\begin{abstract}

%Mass ejections associated with solar flares\cite{Shibata2011,Emslie2012} often affect Earth's environment\cite{Gopalswamy2003}. 
%Active stars sometimes show larger `superflares'\cite{Maehara2012,Notsu2019}, which can cause huge mass ejections, greatly affecting the exoplanet habitability\cite{Airapetian2016,Airapetian2020}. 
%However, the physical properties of stellar mass ejections hitherto have been little known\cite{Moschou2019}, and moreover, no observational indications of such events have been obtained previously especially on solar-type stars. 
%Here we report the first successful optical spectroscopic observation of a large superflare on the solar-type star EK Draconis (EK Dra) that provides conclusive evidence for an associated stellar mass ejection. 
%After the superflare brightened with the radiated energy of $2.0\times10^{33}$ erg, a blue-shifted hydrogen absorption component with a velocity of $-510$ km s$^{-1}$ appeared, with the velocity gradually slowing down over $2$ hours. 
%The temporal changes in the spectra and velocity greatly resemble that of solar mass ejections on the disc. 
%The large ejected mass of $6.2\times10^{17}$ g is consistent with the solar flare-energy/ejected-mass relation. 
%These results demonstrate that a huge and fast stellar mass ejection did occur and that it is similar to solar events. 
%This discovery provides a proxy of how extreme solar and stellar events affect the environments of young/current Earth and (exo)planets, and also can impact stellar mass evolution.

Solar flares are often accompanied by filament/prominence eruptions ($\sim10^{4}$ K and $\sim 10^{10-11}$ cm$^{-3}$), sometimes leading to coronal mass ejections (CMEs) that directly affect the Earth's environment\cite{Shibata2011,Gopalswamy2003}. 
`Superflares' are found on some active solar-type (G-type main-sequence) stars\cite{Maehara2012,Shibata2013,Notsu2019}, but the association of filament eruptions/CMEs has not been established.
Here we show that our optical spectroscopic observation of the young solar-type star EK Draconis reveals the evidence for a stellar filament eruption associated with a superflare.
This superflare emitted a radiated energy of $2.0\times10^{33}$ erg, and blue-shifted hydrogen absorption component with a large velocity of $-510$ km s$^{-1}$ was observed shortly after.
The temporal changes in the spectra greatly resemble those of solar filament eruptions. Comparing this eruption with solar filament eruptions in terms of the length scale and velocity strongly suggests that a stellar CME occurred. 
The erupted filament mass of $1.1\times10^{18}$ g is 10 times larger than those of the largest solar CMEs. 
The massive filament eruption and an associated CME provide the opportunity to evaluate how they affect the environment of young exoplanets/young Earth\cite{Airapetian2020} and stellar mass/angular-momentum evolution\cite{Osten2015}. 

%After the superflare with radiated energy of $2.0\times10^{33}$ erg, a blue-shifted hydrogen absorption component with a large velocity of $-510$ km s$^{-1}$ appeared. 

\end{abstract}

%\Blue{Solar filament eruptions are phenomena that cool ($\sim 10^{4}$ K) and dense ($\sim 10^{10-11}$ cm$^{-3}$) plasma are erupted associated with solar flares releasing the energy of $10^{29}$ to $10^{32}$ erg\cite{Shibata2011}.}
%\Add{Flares, filament eruptions, and CMEs are separate eruptive phenomena on the Sun that are often affiliated, but not always.}
%\Erase{\sout{Active stars including solar-type stars (G-type main sequence stars) are known to sometimes show larger `superflares' releasing the energy more than $10^{33}$ erg, and}} 
%The fast filament eruptions and outer CMEs (and associated high-energy particles) often directly affect Earth's environment\cite{Gopalswamy2003}.}
%, and furthermore, number of events showing clear signatures of filament eruptions/CMEs reported before are \Blue{limited} even for cooler stars and giant stars (cf. descriptions in Methods). 

Solar flares, filament eruptions, and CMEs are thought to be caused by the common magnetohydrodynamic process, though all of them are not necessarily observed in the same event.
Magnetic reconnection is a key energy release mechanism for flares, which are thought to be sometimes be triggered by the instability of cool filaments in active regions\cite{Shibata2011}. 
%They are not necessarily accompanied by all of them, as they each have different observational aspects.
%\Add{These phenomena are not necessarily observed in one event.} %, because they each represent different observational aspects.}
Recently, it has been discussed that much larger ``superflares" that release the energy of more than $10^{33}$ erg (10 times the largest solar flares $\sim10^{32}$ erg) can occur -- or have occurred relatively recently -- even on the Sun\cite{Maehara2012,Notsu2019,Shibata2013,Miyake2012}.
Superflares may produce much larger CMEs than the largest solar flares, which can significantly affect the environment, habitability, and development of life around young and intermediate age stars\cite{Airapetian2020}. 
However, superflares on solar-type stars have been mainly detected by optical photometry (e.g., Kepler space telescope)\cite{Maehara2012}.
Therefore, no observational indication of filament eruptions/CMEs has been reported for solar-type stars.
Optical spectroscopic observations are a promising way to detect stellar filament eruptions, which can be indirect evidence of CMEs.
However, for solar-type stars, optical spectra of superflares have never been obtained.

\begin{figure}
\begin{center}
\includegraphics[width=1.0\linewidth,clip]{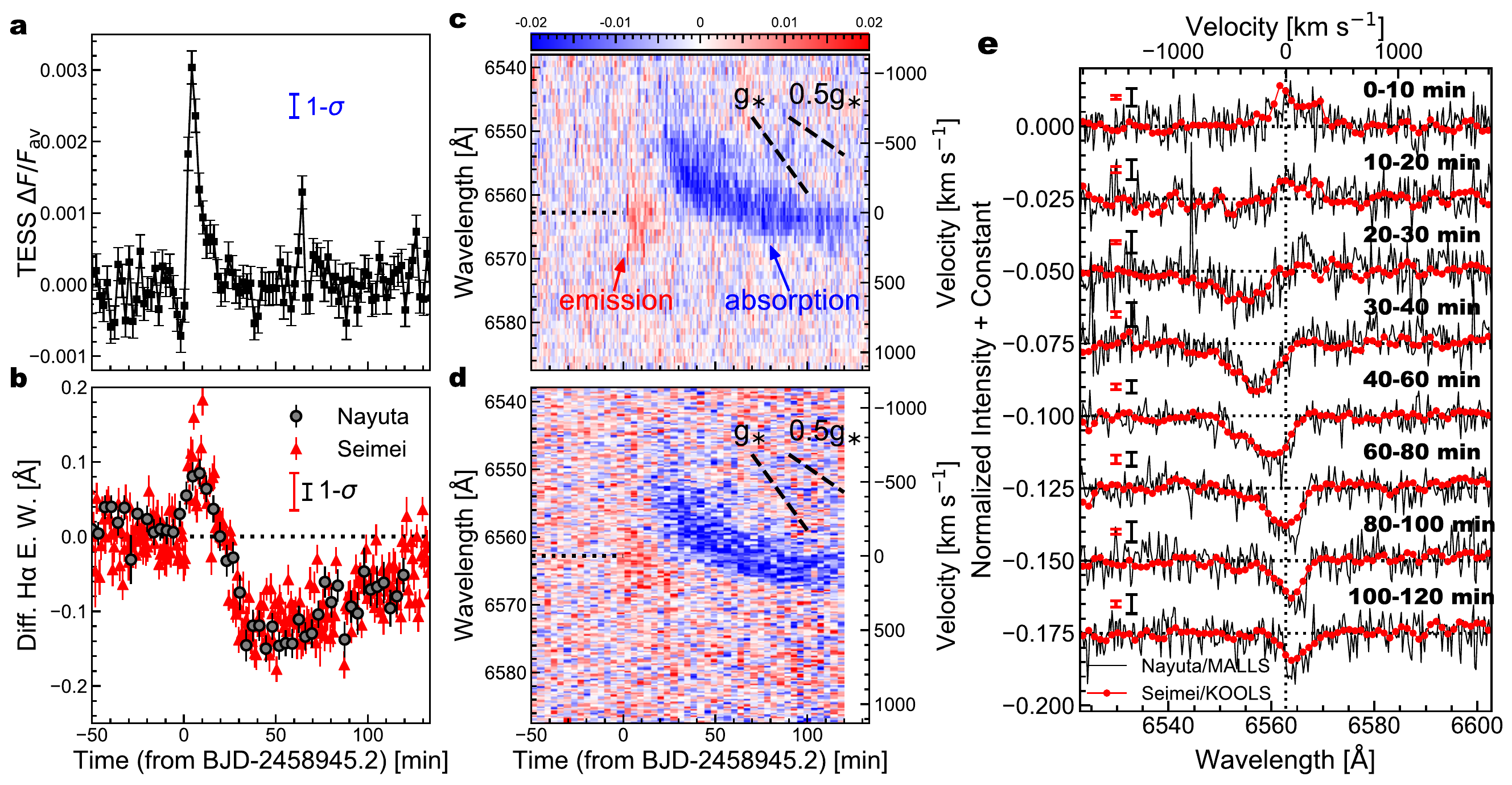}
\caption{ Light curves (a-b) and spectra (c-e) of a superflare on EK Dra. \sf \textbf{(a)} The light curve observed by TESS in white light ($\sim$6000--10,000 {\AA}) on BJD (Barycentric Julian Day) 2458945.2 (5 April 2020). 
The individual points represent the stellar flux normalized by the averaged value with the pre-flare level subtracted. 
The 1-$\sigma$ value of the pre-flare light curve (-150 min to 0 min) is plotted in blue. 
\textbf{(b)} Light curves of the H$\alpha$ equivalent width (E.W.) observed by the medium-dispersion spectroscopy MALLS at the Nayuta telescope (grey circles) and the low-dispersion spectrograph KOOLS-IFU installed at the Seimei telescope (red triangles) during the same observing period as in panel \textbf{(c)}. 
The H$\alpha$ emissions were integrated within $\pm$10 {\AA} from the H$\alpha$ line center (6562.8 {\AA}) after dividing by the continuum level, and the pre-flare level is subtracted. The positive and negative values represent emission and absorption, respectively, compared to the pre-flare level.
The 1-$\sigma$ value of the pre-flare light curve (-150 min to 0 min) is plotted with red and black color for Seimei and Nayuta data, respectively.  
\textbf{(c-d)} Two-dimensional H$\alpha$ spectra obtained by the Seimei Telescope \textbf{(c)} and the Nayuta Telescope \textbf{(d)}. The red and blue colors correspond to emission and absorption, respectively. The dashed lines indicate the stellar surface gravity ($g_{\ast}$) and half of the surface gravity (0.5 $g_{\ast}$). The panels \textbf{(c-d)} share the upper color bar.
\textbf{(e)} Temporal evolution of the pre-flare-subtracted H$\alpha$ spectra observed by the Seimei telescope (red) and the Nayuta telescope (black), with the spectra shifted by constant values for clarity. The spectra are binned in time, and the integration periods correspond to the horizontal axes of panels \textbf{(a-d)}. The intensities are normalized by the stellar continuum level. The vertical dotted line indicates the H$\alpha$ line center, and the horizontal dotted lines indicate the zero levels for each spectrum.  
The 1-$\sigma$ error bar around the line core is also plotted based on the residual scattering in the line wing.
}
\label{Figure1}
\end{center}
\end{figure}

EK Draconis (EK Dra) is known to be an active young solar-type star (a G-type, zero-age main-sequence star with an effective temperature of 5560--5700 K and age of 50--125 million years\cite{Waite2017}) that exhibits frequent UV stellar flares\cite{Ayres2015,Audard1999} and gigantic starspots at low-high latitudes\cite{Waite2017}.
We conducted optical spectroscopic monitoring of EK Dra for 19 nights between 21 January 2020 and 15 April 2020, simultaneously with optical photometry from the Transiting Exoplanet Survey Satellite (TESS)\cite{Ricker2014}.
%Time-resolved neutral-hydrogen H$\alpha$-line spectra at 6562.8 {\AA} (radiation from cool plasma of a few times 10,000 K) were \Blue{observed} using the low-dispersion spectrograph KOOLS-IFU installed at the 3.8-m Seimei Telescope\cite{Kurita2020} and the middle-dispersion spectroscopy MALLS at the 2-m Nayuta Telescope.
Time-resolved neutral-hydrogen H$\alpha$-line spectra at 6562.8 {\AA} (radiation from cool plasma of a few times 10,000 K) were spectroscopically observed at the 3.8-m Seimei Telescope\cite{Kurita2020} and the 2-m Nayuta Telescope.
In this campaign, we succeeded in obtaining optical spectra of large superflares on a solar-type star. 
The superflare that occurred on 5 April 2020 was simultaneously observed using TESS photometry in white light ($\sim$6,000--10,000 {\AA}) and ground-based spectroscopy in H$\alpha$ line (Fig. \ref{Figure1}a-b and Extended Data Fig. 1).
The H$\alpha$ brightening was associated with the TESS white-light flare, which lasted 16 $\pm$ 2 min. 
The radiated bolometric energy of the TESS white-light flare is estimated to be 2.0$\pm$0.1$\times$10$^{33}$ erg (20 times the most energetic solar flares), and the radiated H$\alpha$-line energy was 1.7$\pm$0.1$\times$10$^{31}$ erg; thus, the flare is classified as a superflare.

After the impulsive phase, the TESS white-light intensity returned to its pre-flare level.
However, the equivalent width (hereafter E.W.) of H$\alpha$ (the wavelength-integrated H$\alpha$ emission normalized by the continuum level) became lower than the pre-flare level (i.e., it displayed enhanced absorption), returning to the pre-flare level in approximately 2 hours (Fig. \ref{Figure1}b). 
The blue-shift H$\alpha$ absorption component with a maximum central velocity of about $-$510 km s$^{-1}$ and a half-width of $\pm$220 km s$^{-1}$ appeared soon after the superflare.
The velocity gradually slowed down with time, and a red-shifted absorption component appeared at a few times 10 km s$^{-1}$ (Fig. \ref{Figure1}c-e, Extended Data Fig. 2a, 3a).  
Both ground-based spectroscopic observations simultaneously recorded the same spectral change, demonstrating that low-temperature and high-density neutral plasma above the stellar disk moves at high speed toward the observer before some parts finally start to fall back to the surface. 
In addition, the deceleration is not monotonic: it was 0.34$\pm$0.04 km s$^{-2}$ in the initial phase, dropping to 0.016$\pm$0.008 km s$^{-2}$ in the later phase (Fig. \ref{Figure1}c-d and Extended Data Fig. 3b). 
This is interpreted in terms of changes in the height of the ejected mass. 
The observed deceleration is in good agreement with that due to the surface gravity of approximately 0.30$\pm$0.05 km s$^{-2}$ (ref.\cite{Waite2017}), although the initial value is slightly larger.

\begin{figure}
\begin{center}
\includegraphics[width=11cm]{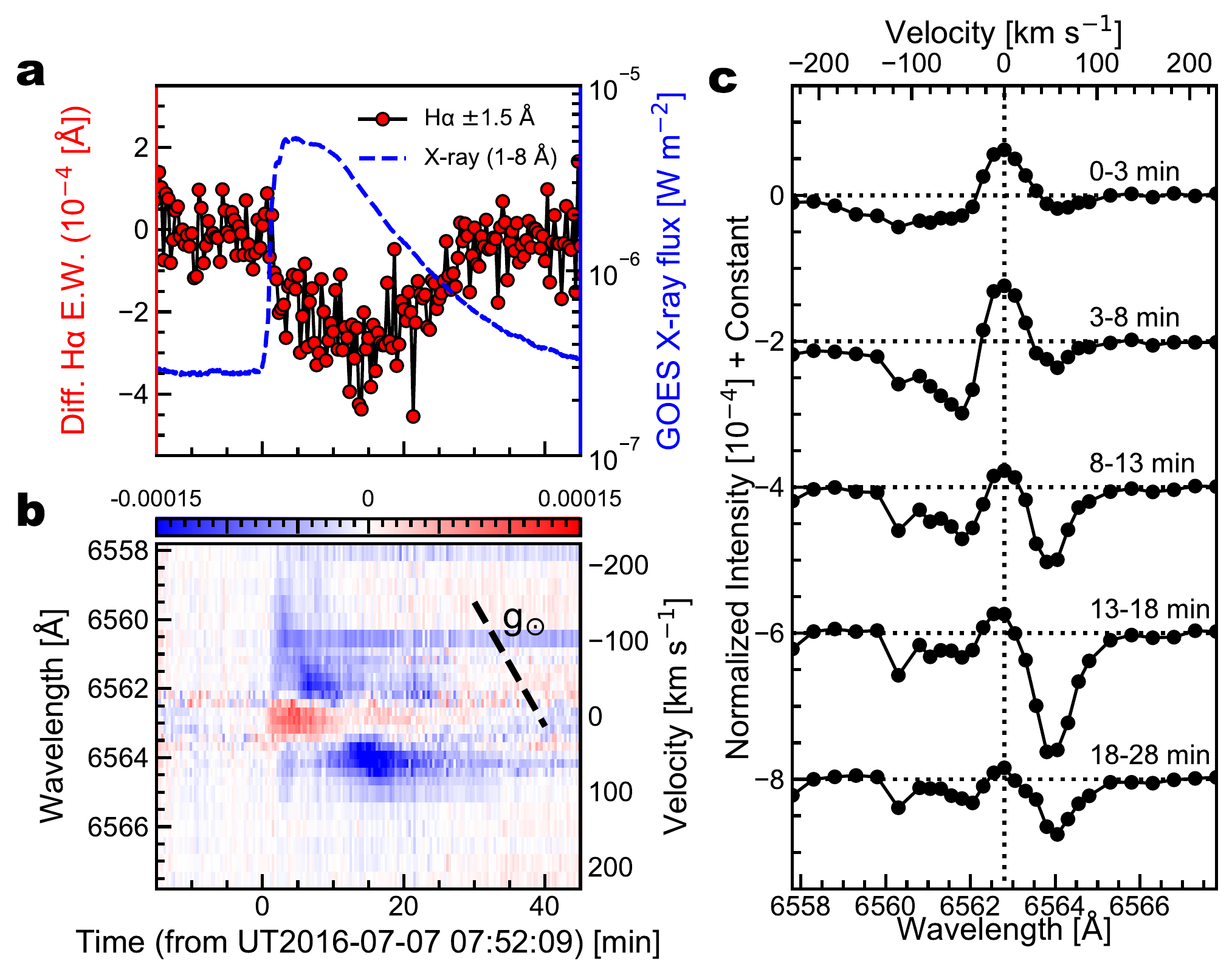}
\caption{
The space-integrated light curves (a) and spectra (b-c) of a C5.1-class solar flare and filament eruption on 7 July, 2016, observed with the SDDI installed at the SMART telescope. \sf \textbf{(a)} GOES soft X-ray (1-8 {\AA}) and H$\alpha$ E.W. light curves of the solar flare are plotted as a blue dashed line and red circles, respectively. 
The H$\alpha$ emissions were integrated within $\pm$1.5 {\AA} from the H$\alpha$ line center (6562.8 {\AA}) and were divided by the full-disk continuum level, and the pre-flare level is subtracted. Time 0 is the time when the flare begins.
\textbf{(b)} Two-dimensional pre-flare-subtracted H$\alpha$ spectra. The red and blue colors correspond to emission and absorption compared to the pre-flare levels, respectively. The dashed line indicates surface gravity at the solar surface. 
\textbf{(c)} 
Temporal evolution of the pre-flare-subtracted H$\alpha$ spectrum shifted vertically by constant values for clarity. The H$\alpha$ spectra were produced by integrating the data over a large enough region to cover the flaring area (see Extended Data Fig. 4).
The intensities are normalized by the total solar continuum level. 
The vertical dotted line indicates the H$\alpha$ line center, and the horizontal dotted lines indicate the zero levels for each spectrum.
}
\label{Figure2}
\end{center}
\end{figure}

How much do the stellar spectral changes obtained here actually resemble those of solar filament eruptions?
Blue-shifted H$\alpha$ absorption profiles are often observed from solar filament eruptions\cite{Schmieder1987,Shibata2011}. 
As in Fig. \ref{Figure2}, we generated spatially-integrated H$\alpha$ spectra of a solar flare/filament eruption that occurred on the solar disk using the SMART data\cite{Ichimoto2017}  (Extended Data Fig. 4, Supplementary Movie 1).
We converted to the full-disk pre-flare subtracted spectra by multiplying by the partial-region/full-disk ratio (i.e., virtual Sun-as-a-star spectra). 
We found that the blue-shifted absorption component at approximately 100 km s$^{-1}$ was predominant soon after the solar flare, and the spatially integrated H$\alpha$ E.W. showed enhanced absorption (Fig. \ref{Figure2}a).
These blue-shifted profiles are unequivocally due to the filament eruption.
Later, the blue-shifted component decelerated and gradually turned into slow, red-shifted absorption (Fig. \ref{Figure2}b-c).
The H$\alpha$ E.W. returned to the pre-flare level in approximately 40 min (Fig. \ref{Figure2}a). 
Although the energy scales and velocities are different, the solar data greatly resembles the spectral changes in the superflare on EK Dra (see Supplementary Information for another event).
This similarity suggests that the stellar phenomenon is the same as the simply magnified picture of the solar filament eruption.

A filament eruption is the only explanation for the blue-shifted absorption component on EK Dra by solar analogy\cite{Shibata2011}.
The hypothesis that the blue-shifted absorption on EK Dra might come from up-/down-flow in flare kernels must be rejected because they never show H$\alpha$ absorption\cite{Svestka1962,Ichimoto1984}. 
Also, down-flow in cooled magnetic loops (known as post-flare loops)\cite{Schmieder1987} show red-shifted absorption, so they cannot explain the blue-shifted absorption. 
(However, the red-shifted absorption in EK Dra in the later phase might be caused by post-flare loops\cite{Schmieder1987}.) 
Rotational visibilities of prominences or spots also are not adequate to explain it since the rotation speed of EK Dra is only 16.4$\pm$0.1 km s$^{-1}$ (ref.\cite{Waite2017}). 
Thus, we concluded that we detected a stellar filament eruption on the solar-type star.

Some observational signatures for stellar filament eruptions or CMEs have been reported previously for cooler K-M dwarfs\cite{Houdebine1990,Honda2018,Leitzinger2011,Vida2016,Veronig2021} and evolved giant stars\cite{Argiroffi2019} (see Methods and ref.\cite{Moschou2019,Airapetian2020} for review).
The observation of a giant star shows a blue-shifted X-ray emission line of 90 km s$^{-1}$ in the post-flare phase and hotter-CME is proposed as a possible explanation\cite{Argiroffi2019}.
Recently, X-ray/EUV dimmings are reported as an indirect evidence of stellar CMEs on K-M dwarfs\cite{Veronig2021}.
In M-dwarf flares, many blue-shifted Balmer/UV line emission components have been reported\cite{Moschou2019,Houdebine1990,Honda2018,Leitzinger2011,Vida2016}, which are interpreted as filament eruptions.  
Some M-dwarf flares share properties similar to the eruption on EK Dra: the blue-shift emissions have high velocities of hundreds of km s$^{-1}$, and some exhibit velocity changes and appear after the impulsive phase\cite{Leitzinger2011,Vida2016}.
For M-dwarf events, the number of studies reporting highly-time-resolved velocity variations of blue-shift components is still insignificant ($\sim$5-min cadence), and simultaneous white-light flare has never been detected.
Our detection of a stellar filament eruption is reliable because we provided solar counterparts,  highly time-resolved spectra ($\sim$50-sec cadence), and simultaneous TESS white-light flare.
%Therefore, this is the most plausible detection of a filament eruption on a main sequence star using observational techniques that are truly analogous to how such eruptions are observed on the Sun, providing more confidence that we observed the same phenomenon.}

What properties does the filament eruption on EK Dra have?
The maximum observed velocity of the blue-shifted component was $\sim-$510 km s$^{-1}$ with a width of 220 km s$^{-1}$. 
This is larger than the typical velocities of solar filament eruptions (10--400 km s$^{-1}$) associated with CMEs\cite{Gopalswamy2003}, although it is a little smaller than the escape velocity at the surface on EK Dra ($\sim$670 km s$^{-1}$).
The cool plasma reached at least $\sim$1.0 stellar radii from the stellar surface (or the initial height) as derived by integrating the velocity over time  (or $\sim$3.2 stellar radii from the stellar surface based on the deceleration rates).
In this case, the projection angle can be allowed at most 45$^\circ$ when we assume the event occurs on the disk center. 
On this projection angle, the velocity can be up to $\sim-$720 km s$^{-1}$, so there is a possibility that the velocities of some components of the EK Dra eruption could exceed the escape velocity.
However, it should be noted that there are weak red-shifted components with a velocity of a few 10 km s$^{-1}$ in the late phase, indicating some materials fell back to the star.
This is often observed in the case of solar filament eruptions with CMEs\cite{Wood2016}.

\begin{figure}
\begin{center}
\includegraphics[width=8cm]{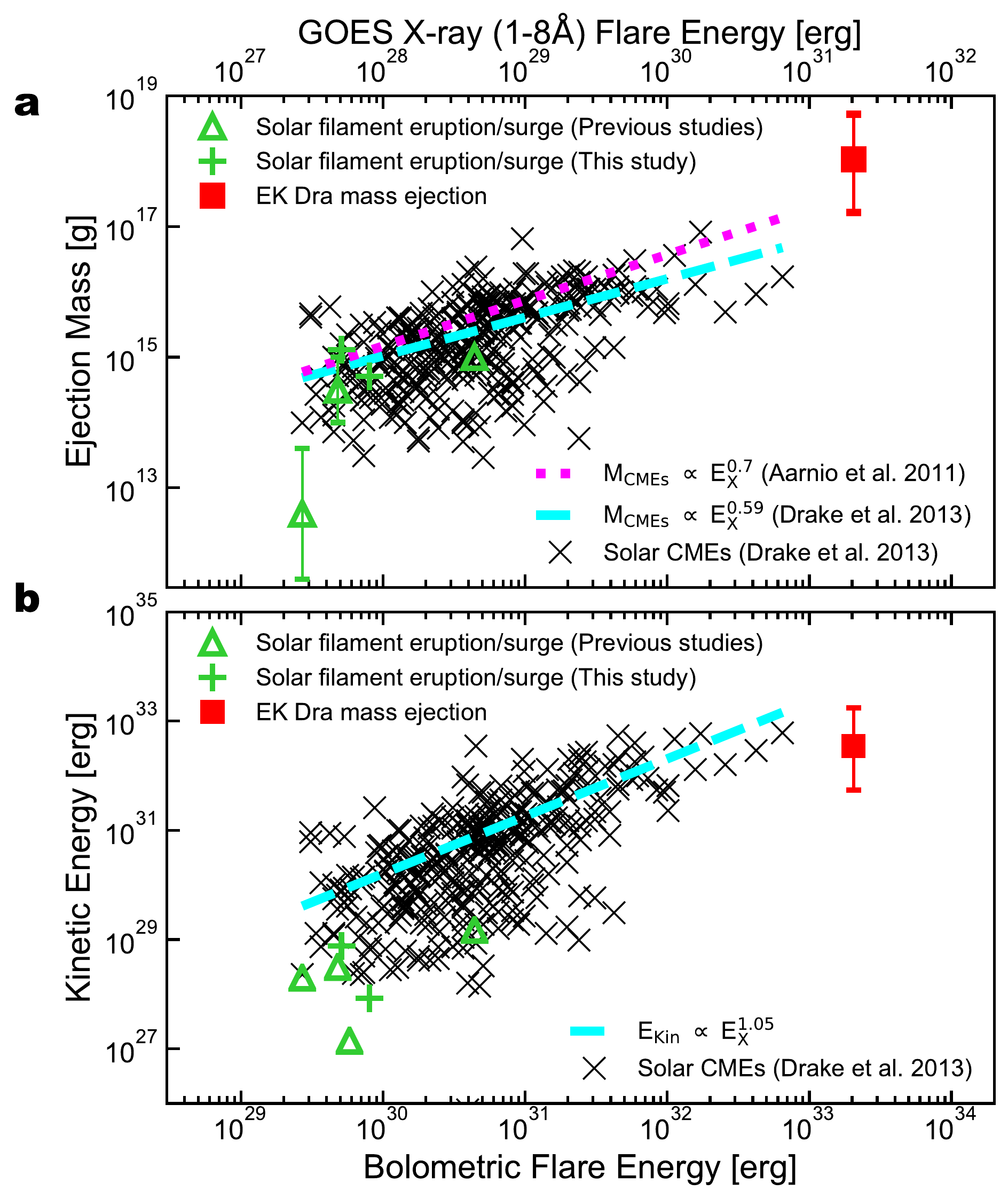}
\end{center}
\caption{
Mass and kinetic energy as a function of flare energy for solar and stellar flares and filament eruptions/CMEs. \sf \textbf{(a)} Comparison between bolometric flare energy and ejected mass. The red square represents the superflare on EK Dra, the black crosses denote for solar CME data, the green triangles are data for solar prominence/filament eruptions and surges taken from previous studies, and the green plus sign signifies the solar filament eruption/surges displayed in Fig. \ref{Figure2} and Supplementary Fig. 9 (Supplementary Information ``Velocity, mass, and kinetic energy: solar data"), respectively (see Table 1).
Note that solar ``surges" are jet-like filament eruption phenomena (see Supplementary Information section ``Another case of solar flares on 2 April 2017" for the explanation of the surge). 
The cyan dashed, and magenta dotted lines are trend fits for solar CMEs expressed as $M_{\rm CMEs} \propto E^{0.59}$ and $M_{\rm CMEs} \propto E^{0.7}$, respectively (see Supplementary Information section ``Solar flare energy-CME mass relation" and and ref.\cite{Aarnio2012,Drake2013}). \textbf{(b)} Comparison between flare bolometric energy and  kinetic energy of the erupted mass. The symbols are the same as in panel \textbf{(a)}. The cyan dashed line is a fit for solar CMEs expressed as $E_{\rm Kin} \propto E_{\rm X}^{1.05}$. 
The kinetic energy of eruption on EK Dra is calculated to be 3.5$^{+14.0}_{-3.0}$$\times$10$^{32}$ erg, which is outside the error range of the predicted value of 4.8$^{+1.1}_{-0.9}$$\times$10$^{33}$ erg; ref.\cite{Drake2013}.
}
\label{Figure3}
\end{figure}

The filament area is estimated to be 1.6$\times 10^{21}$ cm$^2$ (5.6 {\%} of the stellar disk), and the erupted mass is calculated to be 1.1$^{+4.2}_{-0.9}$$\times$10$^{18}$ g based on the absorption components. 
The mass is more than 10 times larger than those of the largest solar CMEs\cite{Aarnio2012,Drake2013} (it should be noted that the mass can be somewhat under-/over-estimated, see Methods).
This mass estimate is in reasonable agreement with those predicted from empirical\cite{Aarnio2012,Drake2013}, and theoretical\cite{Takahashi2016} solar scaling relations between CME mass and flare energy within the error bars ($\sim$ 9.4$^{+3.2}_{-2.4}\times10^{16}$ and 3.1$^{+1.6}_{-1.1}\times10^{17}$ g for ref.\cite{Drake2013} and ref.\cite{Aarnio2012}, respectively) (Fig. \ref{Figure3}a). 
This suggests that the stellar filament eruption can share the common underlying mechanism with smaller-scale filament eruptions/CMEs (i.e., magnetic energy release\cite{Shibata2011,Takahashi2016}) although the absolute values of most physical quantities are very different.

Moreover, the kinetic energy is calculated to be 3.5$^{+14.0}_{-3.0}$$\times$10$^{32}$ erg, which is 16 {\%} of radiation energy in white light. 
The magnetic energy stored around the starspots on EK Dra can be at least 8.0$\times$10$^{35}$ erg, which is enough to produce superflares and filament eruptions with energy of $\sim$10$^{33}$ erg.
In addition, this value is slightly smaller than those extrapolated from the solar CME scaling law (4.8$^{+1.1}_{-0.9}$$\times$10$^{33}$ erg; ref.\cite{Drake2013}) (Fig. \ref{Figure3}b), which is similar to the filament eruption/CME candidates on other stars\cite{Moschou2019}.
In previous studies, it has been argued that  kinetic energy can be reduced by overlaying magnetic fields\cite{Alvarado-Gomez2018,Moschou2019}.
The deceleration of our events was a few times 10 {\%} larger than the stellar gravity (Extended Data Fig. 3b). The strong magnetic fields on EK Dra were reported before\cite{Waite2017} and may support the above explanations.
However, its small kinetic energy can also be understood through a solar analogy: The velocities of (lower-lying) filament eruptions are usually 4-8 times slower than those of the corresponding (higher-lying) CMEs\cite{Gopalswamy2003}, and therefore the kinetic energies of filament eruptions are typically smaller (green symbols in Fig. \ref{Figure3}b).

\begin{figure}
\begin{center}
\includegraphics[width=8cm]{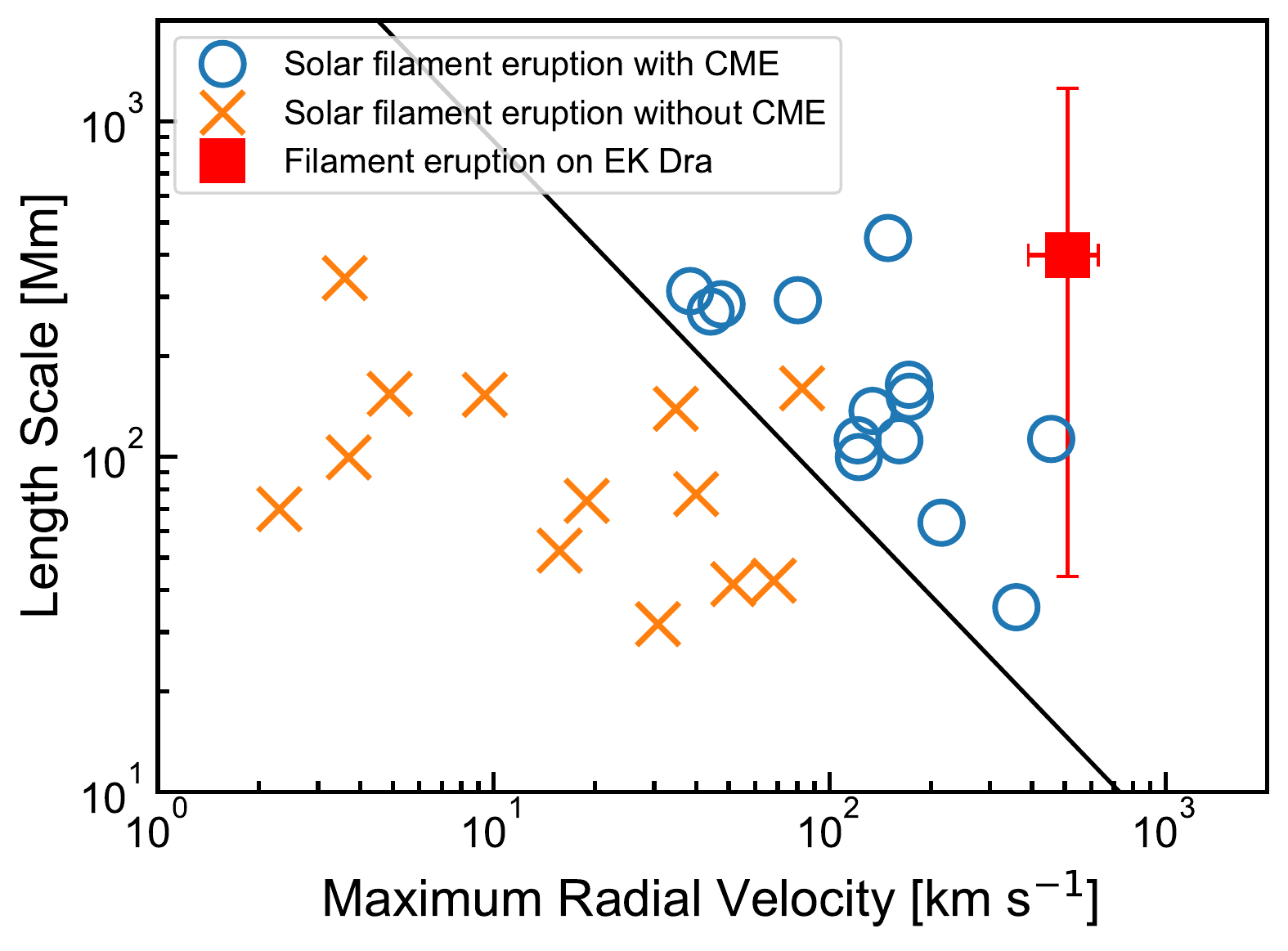}
\end{center}
\caption{
Statistical properties of solar filament eruptions taken from ref.\cite{Seki2021} and comparison with the stellar filament eruption on EK Dra. \sf 
The figure shows the comparison between maximum radial velocities ($V_{\rm r\_max}$) and length scales ($L$) of filament eruptions. 
The blue circles and orange crosses indicate the solar filament eruptions with and without CMEs, respectively.
The red point corresponds to the stellar filament eruption on EK Dra obtained here (see Methods section ``Velocity, mass, and kinetic energy: stellar data" for the calculation of the length scale).
Note that the stellar data are the observed line-of-sight velocity.
Since the deceleration of the stellar filament eruption corresponds well with the surface gravity and the absorption component of filament is visible on the disk all time, we can expect that the filament is flying in our direction perpendicularly to some extent, so there would not be such a big difference between radial velocity and line-of-sight velocity.
We expect that the radial velocity can be larger than the line-of-sight velocity if we assume the projection effect, while it will be about $\sqrt{2}$ times smaller at most if it erupts at a 45-degree tilt in the radial direction, which does not change our discussion.
The solid line indicates the threshold that can roughly distinguish filament eruption with and without CMEs derived by ref.\cite{Seki2021}.
The threshold can be expressed as $(V_{\rm r\_max}/100$ km s$^{-1}$) $(L / 100$ Mm)$^{0.96} = 0.8$, which is determined by using the algorithm of Linear Support Vector Classification (see ref.\cite{Seki2021} for the detailed method).
}
\label{FigS0}
\end{figure}

Did a CME occur in this event? 
Obviously, the line-of-sight velocity $\sim$510 km s$^{-1}$ was slower than the escape velocity and some masses fell back, which may indicate a so-called ``failed" filament eruption\cite{Alvarado-Gomez2018}.
However, this does not necessarily mean that a CME did not occur, again by solar analogy.
In fact, the erupted filaments often fall back to the Sun while CMEs happen.
For example, a well-studied solar event on 2011 June 7 involved a 200-600 km/s filament eruption where lots of filamentary material fell back to the Sun, but some mass clearly escapes as a CME with velocities of $\sim$1000 km s$^{-1}$  (see ref.\cite{Wood2016} and Supplementary Information). 
The event on EK Dra may correspond to this solar event.
In addition, ref.\cite{Seki2021} showed that whether a solar filament eruption leads to a CME can be simply distinguished by a parameter of $(V_{\rm r\_max}/100$ ${\rm km}$ s$^{-1}$) $(L / 100$ ${\rm Mm})^{0.96}$, where $V_{\rm r\_max}$ is the maximum radial velocity and $L$ is the length scale (Fig. \ref{FigS0}).
When the parameter is more than $\sim 0.8$, the probability that a filament eruption lead to a CME is more than 90\%\cite{Seki2021}.
The value of the parameter of eruption on EK Dra is $\sim$18, meaning that our detection of the fast and sizeable stellar filament eruption is indirect evidence that mass escapes into interplanetary space as a CME.

Finally, we summarize future directions of our findings (see Supplementary Information for details):
It is speculated that the filament eruptions/CMEs associated with superflares can severely affect planetary atmospheres\cite{Airapetian2020}.
Our findings can therefore provide a proxy for the possible enormous filament eruptions on young solar-type stars and the Sun, which would enable us to evaluate the effects on the ancient, young Solar-System planets and the Earth, respectively. 
Further, it is also speculated that stellar mass loss due to filament eruptions/CMEs can more significantly affect the evolutionary theory of stellar mass, angular momentum, and luminosity\cite{Osten2015,Aarnio2012}, than stellar winds.
At present, frequency and statical properties of CMEs on solar-type star is unknown, but important insights into these points are obtained by increasing the samples in the future.

\clearpage

\renewcommand{\figurename}{\scriptsize \bf\sffamily\noindent{Fig.}}

\renewcommand{\figurename}{\scriptsize \bf\sffamily\noindent{Table}}

\begin{table}
 \label{table:solarmass}
 \begin{threeparttable}
 \centering
  \begin{tabular}{clcccll}
   \hline
   Date & GOES & flare energy & mass & kinetic energy & reference & event \\
   yyyy/mm/dd & & [10$^{29}$ erg] & [g] & [erg]  & & \\
   \hline \hline
   1980/10/30 & C4.8 & 4.5 & 10$^{14-15}$ & 3.14$\times 10^{28}$ & Ref.\cite{Jain1987} & surge\\
   2001/8/30 & C5.8 & 5.5 & - & 1.4$\times 10^{27}$ & Ref.\cite{Liu2004} & surge\\
   1993/5/14 & M4.4 & 44 & 10$^{15}$ & 1.5$\times 10^{29}$ & Ref.\cite{Ohyama1999} & filament eruption\\
   2012/2/11 & C2.7 & 2.7 & 4$\times$10$^{11-13}$ & 2$\times 10^{28}$ & Ref.\cite{Christian2015} & filament eruption\\
   2016/7/7 & C5.1 & 5.1 & 1.3$\times 10^{15}$ & 7.7$\times 10^{28}$ & Event-1 & filament eruption \\
   2017/4/2 & C8.0 & 8.0 & 5.1$\times 10^{14}$ & 8.5$\times 10^{27}$ & Event-2$^{\dagger}$ & surge\\
   \hline
  \end{tabular}
  \begin{tablenotes}
	\item[$\dagger$]{The analysis of the solar surge (Event-2) is described in Supplementary Information section ``Another case of solar flares on 2 April 2017".}
\end{tablenotes}
\end{threeparttable}
\caption{Properties of solar filament eruptions/surges reported in previous studies and this study. The data are plotted in Figure \ref{Figure3}. The calculation of flare energy, mass, and kinetic energy is introduced in Supplementary Information section ``Velocity, mass, and kinetic energy: solar data".}
\end{table}

\clearpage

\begin{methods}

\subsection{TESS light-curve analysis}

TESS observed EK Dra (TIC 159613900) in its sector 14-16 (18 July 2019-6 October 2020) and 21-23 (21 January 2020-15 April 2020). 
The TESS light curve from the 2-min time-cadence photometry was processed by the Science Processing Operations Center pipeline, a descendant of the Kepler mission pipeline based at the NASA Ames Research Center\cite{Ricker2014,Fausnaugh2020}. 
Extended Data Fig. 1 shows the light curve of the EK Dra from BJD (Barycentric Julian Day) 2458945 (= JD 2458944.997 = 5 April 2020 11:56UT; Sector 23), and the stellar superflare detected by TESS, Seimei telescope, and Nayuta telescope in Fig. \ref{Figure1} is indicated with the red arrow in this figure. 
The quasi-periodic brightness variation is thought to be caused by the rotation of EK Dra with the asymmetrically-spotted hemisphere\cite{Maehara2012,Notsu2019}.
The rotation period is reported as about 2.8 days\cite{Waite2017}.
Although the superflare occurred near the local brightness maximum, some of the starspots are expected to be visible from the observer\cite{Roettenbacher2018,Doyle2018,Doyle2020,Namekata2020a}.
In Extended Data Fig. 1, other flares are also indicated with black arrows with more than two 
consecutive observational points whose flaring amplitude is more than 3 times TESS photometric errors\cite{Maehara2012,Shibayama2013}. 
The white-light flare energy was calculated by assuming the 10,000 K blackbody spectra\cite{Shibayama2013,Namekata2017} (see, Supplementary Information section ``Flare energy").
The pixel-level data analysis is shown in Supplementary Information section ``TESS pixel-level data analysis".
The estimated flare occurrence frequency of superflares ($>$ 10$^{33}$ erg) in the TESS band was about once per 2 days, which means that  about-twelve-nights monitoring observations are necessary on average to detect one superflare from the ground-based telescope under the clear-sky ratio of 50 \%.
This implies that our datasets are highly unique.

\subsection{Spectroscopic data analysis}
 
Here, we present the utilization of low-resolution spectroscopic data from KOOLS-IFU\cite{Matsubayashi2019} (Kyoto Okayama Optical Low-dispersion Spectrograph with optical-fiber Integral Field Unit) of the 3.8-m Seimei Telescope\cite{Kurita2020} at Okayama Observatory of Kyoto University and MALLS\cite{Ozaki2005,Honda2018} (Medium And Low-dispersion Long-slit Spectrograph) of the 2-m Nayuta Telescope at Nishi-Harima Astronomical Observatory of University of Hyogo. 
KOOLS-IFU is an optical spectrograph with a spectral resolution of R ($\lambda/\Delta \lambda$) $\sim$ 2,000 covering a wavelength range from 5800 to 8000 {\AA}; it is equipped with Ne gas emission lines for wavelength calibration and instrument characterization.
The exposure time was set to be 30 sec for this night. 
The sky spectrum was subtracted by using the sky fibers for each spectrum.
The data reduction follows the prescription in ref.\cite{Namekata2020}.
During this observation, the signal-to-noise ratio (S/N) for one frame is typically 172$\pm$6.
The observations by Seimei Telescope ended just after 133.7 min in Fig. \ref{Figure1}b-d. 

MALLS is optical spectroscopy with a spectral resolution of R $\sim$ 10,000 at the H$\alpha$ line covering a wavelength range from 6350 to 6800 {\AA}; it is also equipped with Fe, Ne, and Ar gas emission lines for wavelength calibration and instrument characterization. 
The sky spectrum was subtracted using a nearby region along the slit direction for each observation.
The exposure time was set to be 3 min for this night.
The MALLS data reduction follows the prescription in ref.\cite{Honda2018}. 
The signal-to-noise ratio (S/N) for one frame is typically 86$\pm$8 during this observation.
For the MALLS data, the wavelength corrections are also performed for each spectrum by using the Earth's atmospheric absorption lines. 

We corrected the wavelength for the proper motion velocity of $-$20.7 km s$^{-1}$ of EK Dra based on Gaia Data Release 2 (ref.\cite{Gaia2018}).
Continuum levels are defined by fitting with the linear line between the wavelength range of H$\alpha$ line wing (6517.8-6537.8 and 6587.8-6607.8 {\AA}). 
We take the continuum level as the wavelength range of 6517.8-6537.8 and 6587.8-6607.8 {\AA} to measure the E.W. (= $\int(1-F_{\lambda}/F_{0})d\lambda$, where $F_{0}$ is the continuum intensity on either side of the absorption feature, while $F_{\lambda }$ represents the intensity across the entire wavelength range of interest).
The original spectra are shown in Supplementary Information section ``Stability of pre-flare spectra".
Extended Data Fig. 2 shows the pre-flare-subtracted H$\alpha$ spectra during and after the superflare on EK Dra with higher time cadence than Fig. \ref{Figure1}e.
The narrow-band H$\alpha$ E.W. (H$\alpha$ - 10 {\AA} $\sim$ H$\alpha$ + 10 {\AA}) is used for the measurements of the radiated energy and duration of H$\alpha$ flare because of the high S/N, and the broad-band H$\alpha$ E.W. (H$\alpha$ - 20 {\AA} $\sim$ H$\alpha$ + 10 {\AA}) is used for the measurements of the amount of absorption (i.e., mass and kinetic energy).

\subsection{Solar data analysis}

In the main text, we showed the data of a C5.1-class solar flare (i.e., the peak GOES soft X-ray flux $F_{\rm GOES}$ is 5.1$\times$10$^{-6}$ W m$^{-2}$, hereafter ``Event-1") and associated filament eruption around 07:56 UT, 7 July 2016 observed by the SDDI (Solar Dynamics Doppler Imager)\cite{Ichimoto2017} installed on the SMART (the Solar Magnetic Activity Research Telescope) at Hida Observatory (further explanations continue to the Supplementary Information ``Solar data analysis (continued from Methods)").
The SDDI conducted a monitoring observation of the Sun on 7 July 2016. 
It takes full-disk solar images at 41 wavelength points at every 0.5 {\AA} from the H$\alpha$ line center $-$9.0 {\AA} ($-$411 km s$^{-1}$) to the H$\alpha$ line center +8.0 {\AA} (411 km s$^{-1}$), while it takes the images at every 0.25 {\AA} from the H$\alpha$ line center -2.0 {\AA} (-91 km s$^{-1}$) to the H$\alpha$ line center +2.0 {\AA}  (91 km s$^{-1}$).
Each set of images is obtained with a time cadence of 20 seconds and a pixel size of about 1.2 arcsec. 
The SDDI started the daily monitoring observations in 2016, and the C5.1-class solar flare is one of the largest solar flares with a filament eruption among the events observed by SDDI with good weather conditions in these 5 years.
The solar filament eruption was also reported in ref.\cite{Ichimoto2017,Seki2019}.
Another jet-like filament eruption (known as solar ``surge"\cite{Shibata2011}) associated with a C8.0-class solar flare is also shown in the Supplementary Information section ``Another case of solar flares on 2 April 2017" (hereafter we call this surge ``Event-2").

This paper used 70-min time series of the SDDI images taken from 07:30 UT on 7 July 2016 (see, Supplementary Movie 1). 
As in Extended Data Fig. 4, the C5.1-class flare occurred around an active region, named ``NOAA 12561", on the solar disk and was accompanied by a typical filament eruption\cite{Ichimoto2017,Seki2019}.
The spectra from the event are integrated over a spatial region that is large enough to cover the visible phenomena (the magenta region in Extended Data Fig. 4a-b).
The spectra are reconstructed by using the template solar H$\alpha$ spectrum convolved with SDDI instrumental profile. 

Here, we define $L(\lambda, t, {\rm A})$ as a luminosity at a wavelength of $\lambda$ and time of $t$ which is integrated for the region $A$ (i.e., $L(\lambda, t, {\rm A})$ = $\int_{\rm A}$ $I(t)$ d$A$, $I(t)$ is intensity). 
We now define $A_{\rm local}$ as the integration region (magenta region in Extended Data Fig. 4a-b), and $A_{\rm full-disk}$ as the solar full disk.
We first obtained the local (partial-image) pre-flare subtracted spectra $\Delta S_{\rm local}$ which are normalized by local (partial-image) total continuum level ($L(6570.8 {\rm \AA}, t, A_{\rm local})$):
\begin{eqnarray}
\Delta S_{\rm local} = \frac{L(\lambda, t, A_{\rm local}) - L(\lambda, t_0 , A_{\rm local})}{L(6570.8 {\rm \AA}, t, A_{\rm local})},
\end{eqnarray}
where $t_{0}$ is a given time of the pre-flare period.
Then, the (virtual) full-disk pre-flare-subtracted spectra $\Delta S_{\rm full-disk}$ are obtained by multiplying the ratio of the partial-image continuum to full-disk continuum (total continuum ratio: 
\begin{eqnarray}
\Delta S_{\rm full-disk} = \Delta S_{\rm local} \times \frac{ L(6571.8 {\rm \AA}, t_0, A_{\rm local}) }{ L(6570.8 {\rm \AA}, t_0, A_{\rm full-disk}) } ,
\end{eqnarray}
then we obtained a virtual pre-flare-subtracted spectrum of this phenomena as if we observed the Sun as a star.

The E.W. of the H$\alpha$ is also calculated by using the full-disk-normalized and pre-flare-subtracted spectra ($\Delta S_{\rm full-disk}$), and we obtained the virtual Sun-as-a-star $\Delta$H$\alpha$ E.W. (i.e., differential H$\alpha$ flux normalized by the full-disk continuum level).

\subsection{Velocity, mass, and kinetic energy: stellar data}

For the stellar filament eruption, the velocity is derived by fitting the absorption spectra obtained by Seimei telescope with the normal distribution $N(\lambda, \mu, \sigma^2)$ where $\mu$ is the mean wavelength and $\sigma^2$ is the variance. 
In Extended Data Fig. 3a, we plotted the temporal evolution of the velocity (($\mu-\lambda)/\lambda\times c$, where $\lambda$ is 6562.8 {\AA}, $c$ is light speed) for the fitted absorption feature with the width of $\sigma$. 
We only plotted the data whose absorption features are clear enough to fit the shape with the threshold of the fitted absorption amplitude $>$ 0.01 and fitted velocity dispersion of $<$ 500 km s$^{-1}$ and $>$ 100 km s$^{-1}$. The threshold was determined by trial and error, and we find that many missed detections of absorption features occur when we select threshold values other than this one. The amplitude value of 0.01 corresponds to the detection limit when considering the typical S/N$\sim$170 of the Seimei Telescope/KOOLS-IFU, and the lower limit of 100 km s$^{-1}$ is determined to avoid detecting the sharp noisy signals. About 27\% of data points were discarded due to this threshold from the initial points (22 min) to final points (110 min), especially for the latter decaying phase. 
Here, the maximum observed velocity and its errors are calculated as 510$\pm$120 km s$^{-1}$ with its width of  220$\pm$90 km s$^{-1}$ from the mean values of the $\mu$ and $\sigma$ of the first five points (t = 22-26 min in Fig. \ref{Figure1}), respectively.
The mean values of the velocity when the absorption becomes strong (t = 25-50 min in Fig. \ref{Figure1}) is estimated as 258 km s$^{-1}$.

The plasma mass is simply calculated from the total H$\alpha$ E.W.. 
We used the simple Becker's cloud model\cite{Mein1988} with optical depth at the line center of the ejected plasma  $\tau_0$ of 5 (which is slightly more optically thick than solar filament eruptions; c.f., ref.\cite{Odert2020}), the two-dimensional aspect ratio of 1 (i.e., cubic), local plasma dispersion velocity $W$ of 20 km s$^{-1}$, and source function $S$ of 0.1 based on the solar observations\cite{Sakaue2018}.
The observed half width of 220 km s$^{-1}$ of the stellar blue shifted component is larger by one order of magnitude than the solar value, but here we use the solar value as a template. The dispersion velocity of 220 km s$^{-1}$ is considered to be the upper limit of the local velocity dispersion because the ejected mass would have the complex two-dimensional velocity distribution which can cause larger $W$ in the integrated spectra.
First, modeled E.W. of enhanced absorption is calculated by using the Becker's cloud model when the plasma velocity $v_{\rm shift}$ is $-$258 km s$^{-1}$ as
 \begin{eqnarray}
  {\rm model \,\,  E.W.} &=& \int_{\lambda}\frac{I_{\lambda}- I_{0\lambda}}{I_{0,\rm Cont.}} d\lambda
  = \int_{\lambda}\frac{S-I_{0\lambda}}{I_{0,\rm Cont.}} \left( 1- e^{-\tau_{\lambda}} \right)d\lambda \label{eq:CM1} \\
  \tau_{\lambda} &=& \tau_{0} {\rm exp} \left[ - \frac{1}{2} \left(  \frac{\lambda/\lambda_0 - (1 + v_{\rm shift}/c)}{W/c} \right)^2 \right],
  \label{eq:CM2}
 \end{eqnarray}
where $I_{0\lambda}$ is background intensity and  $I_{0,\rm Cont.}$ is continuum intensity.
This is the E.W. value for an extreme case when the full disk of the star is completely covered with absorbing, cool ejected plasma. 
By comparing the modeled E.W. (Eq. \ref{eq:CM1}) with the lowest observed stellar E.W. value of $-$0.16 {\AA} (integrated for H$\alpha$ - 20 {\AA} $\sim$ H$\alpha$ + 10 {\AA}; see Supplementary Fig. 4c), the cool-plasma filling factor compared to the stellar disk is calculated to be 5.9 \% of stellar disk (i.e., modeled E.W./observed E.W.; $Area$ = 1.6$\times$10$^{21}$ cm$^2$). 
Using the length scale of the ejected plasma 3.9$\times$10$^{10}$ cm (= $Area^{0.5}$), the hydrogen column density is derived  as 4.0$\times10^{20}$ cm$^{-2}$  from the assumed optical depth based on the plasma model\cite{Tsiropoula1997}.
In the model of ref.\cite{Tsiropoula1997}, hydrogen/electron density is calculated by assuming an ionization equilibrium for a population of hydrogen atoms due to a balance between recombination and radiative photoionization through Balmer/Lyman continuum. 
It should be noted that the ionization equilibrium of filaments on active stars may be somewhat different from the solar observations due to their high UV radiations, which may affect the evaluation of the mass of the ejecta.
By multiplying the hydrogen column density by the filament area, we then obtained the plasma mass of 1.1$\times$10$^{18}$ g. 
If the two-dimensional aspect ratio becomes 0.1 similar to jet-like feature (x-width:y-width:z-depth = 1:0.1:0.1), then the estimated mass becomes larger by a factor of 1.78.
If optical depth ranges from 0.8 to 10 (ref.\cite{Odert2020}), the source function takes values of 0.02 or 0.5, and the dispersion velocity takes 10 or 220 km s$^{-1}$\cite{Sakaue2018}, the estimated masses change by a factor of from 0.15 to 4.9.
In Fig. \ref{Figure3}a, we used the mass of 1.1$^{+4.2}_{-0.9}$$\times$10$^{18}$ g for optical depth of 5, and uncertainties of the model (0.15-4.9) are used as the error bars since the model-based errors are expected to be much larger than the observational errors.
It should be noted that this mass estimate could be either a significant overestimate of the mass of an affiliated CME due to most of the filament falling back to the star, or it could be a significant underestimate due to most of the CME actually being hot coronal material rather than cool filament.
The plasma kinetic energy is then calculated as 3.5$^{+14.0}_{-3.0}$$\times$10$^{32}$ erg by using the velocity of 258 km s$^{-1}$. 
The observed maximum velocity was 510 km s$^{-1}$ in the early phase, so the kinetic energy can be larger by a factor of 4 although the absorption component was weak at that time.

\subsection{Related works on candidates of stellar filament eruptions/CMEs on other types of stars}

Here, we discuss potential stellar filament eruptions/CMEs reported in the previous studies (see\cite{Osten2017,Moschou2019,Airapetian2020} for review). 
In other stars, such as M-type stars\cite{Houdebine1990,Gunn1994,Fuhrmeister2004,Fuhrmeister2008,Leitzinger2011,Vida2016,Honda2018,Leitzinger2014,Vida2016,Korhonen2017,Vida2019,Muheki2020,Maehara2020,Zic2020}, K-type stars\cite{Flores-Soriano2017}, T-Tauri stars\cite{Guenther1997,Tsuboi1998}, close binaries\cite{Favata1999,Bond2001}, and giant stars\cite{Argiroffi2019}, some observational candidates of stellar filament eruptions/CMEs have been reported, although confirmations of filament eruptions/CMEs in analogy with solar observations are still rare.
We also note that some other studies have tried to detect a signature of stellar filament eruptions/CMEs in various ways but have not succeeded in robust detection\cite{Leitzinger2010,Boiko2012,Villadsen2016,Crosley2016,Leitzinger2020,Crosley2018a,Crosley2018b,Villadsen2019}.

A signature of CME was reported from a blue-shifted emission component of the cool X-ray O VIII line (4 MK) in the late phase of stellar flare on an evolved giant star HR 9024\cite{Argiroffi2019}.
Although the time evolution of the blue-shifted velocity is not obtained there, they detected the blue-shifted emission component with a velocity of 90 km s$^{-1}$ (the escape velocity 220 km s$^{-1}$) and interpreted it as a CME.
The blue-shifted plasma components with a few MK are also emitted from the upward flow in the confined flare loops (called ``chromospheric evaporation") in the case of solar flares, but they exclude the possibility considering that the other hotter lines do not show the blue-shifted component in the post-flare phase.
Although the spectral-type of HD 9024 (evolved giant star) is very different from EK Dra and the velocity (90 km s$^{-1}$) is smaller than our observation (510 km s$^{-1}$), both observations share the same trend that mass ejection signatures is dominant in the post-flare phase.

Blue-shifted emission components of chromospheric lines have been reported in association with the Balmer-line flares mostly on active M/K dwarfs\cite{Houdebine1990,Gunn1994,Fuhrmeister2004,Fuhrmeister2008,Leitzinger2011,Honda2018,Leitzinger2014,Vida2016,Korhonen2017,Vida2019,Muheki2020,Maehara2020,Zic2020,Flores-Soriano2017} (see ref.\cite{Moschou2019,Odert2020} for the summary).
The time-varying blue-shifted hydrogen emission components have also been reported with high time cadence on M dwarfs (e.g., ref.\cite{Vida2016,Honda2018}). 
A similar case is reported for a UV flare on an M dwarf\cite{Ambruster1986,Leitzinger2011}.
These can be possible evidence of stellar prominence eruptions/CMEs.
It seems quite possible that the blue-shifted emission lines on M-dwarfs are very analogous to the H$\alpha$ absorption signatures studied in this paper.
The fundamental differences between G-dwarf and M-dwarf blue-shift signature is that for hotter G dwarfs, H$\alpha$ in an erupting filament will only be detectable in absorption, whereas for the cooler M dwarfs even the quiescent H$\alpha$ line is in emission, so an erupting filament might be observed in emission as well (cf. ref.\cite{Odert2020}).
Blue-wing enhancements of M-dwarf flares is characterized by the high velocity of several hundred km s$^{-1}$ (sometimes more than that)\cite{Maehara2020,Vida2019,Vida2019,Houdebine1990}, which cannot be explained by  chromospheric-line blue-shift phenomena associated chromospheric evaporation flow observed in solar flares\cite{Svestka1962,Allred2005,Tei2018,Kuridze2015,Odert2020,Heinzel1994}. 
The high velocity of M-dwarf flares are similar to that detected on EK Dra in this study ($\sim$510 km s$^{-1}$).
In addition, not all but some of the blue-shift events on M dwarfs appear after the impulsive phase\cite{Leitzinger2011,Vida2016}, which shares the same properties with filament eruption events on EK Dra and the Sun in this study.
Therefore, at present the blue-shifted emission lines in M-type stars are most likely prominence eruptions.

Other signatures of kinematic characteristics of the ejected plasma are also inferred from continuous X-ray absorption during stellar flares, which can be caused by neutral material above the flaring region, such as filament eruptions\cite{Haisch1983,Ottmann1996,Wheatley1998,Tsuboi1998,Favata1999,Franciosini2001,Pandey2012,Moschou2019}.
However, on the Sun, X-ray absorption by prominences is uncommon\cite{Schwartz2015,Odert2020}, and instrumental calibration effects at low energy have been pointed out\cite{Osten2017}.

In some cases of binary stars, eclipses of the white dwarf component have been interpreted as obscuration by stellar mass ejected from the late-type companion star\cite{Bond2001,Parsons2013}.
Other than this, pre-flare dips have been reported in stellar flares, suggesting potential prominence eruptions/CMEs\cite{Giampapa1982,Doyle1990}.
Radio observations have recently investigated the type-II radio bursts associated with shocks in front of CMEs as possible indirect evidence of CMEs, but no significant signature has been obtained so far\cite{Leitzinger2010,Boiko2012,Villadsen2016,Crosley2016,Osten2017,Crosley2018a,Crosley2018b,Villadsen2019}.
Recently, a stellar type-IV burst event from the M-type star Proxima Centauri was reported and may be evidence for a stellar CME\cite{Zic2020}.

\subsection{Data availability}

In addition to the figure data available, all raw spectroscopic data are available either in the associated observatory archive ({\footnotesize \url{https://smoka.nao.ac.jp/index.jsp}} for KOOLS-IFU data in Figure \ref{Figure1} (available after Jan 2022); {\footnotesize \url{https://www.hida.kyoto-u.ac.jp/SMART/T1.html}} for a part of SDDI data in Figure \ref{Figure2}) or upon request from the corresponding author (for MALLS data in Figure \ref{Figure1} and full raw data of SDDI). The TESS light curve is available at the MAST archive ({\footnotesize \url{https://mast.stsci.edu/portal/Mashup/Clients/Mast/Portal.html}}). All datasets used to make figures are available online.

\end{methods}

%% Here is the endmatter stuff: Supplementary Info, etc.
%% Use \item's to separate, default label is "Acknowledgements"

\begin{addendum}
\item 
Some of the data presented here were obtained at the Okayama Observatory of Kyoto University, which is operated as
a scientific partnership with National Astronomical Observatory of Japan. 
We are grateful to the staff of Okayama Observatory, M. Kurita, M. Kino, F. Iwamuro, K. Ohta, H. Izumiura, K. Matsubayashi, D. Kuroda, and T. Nagata, for the telescope and instrument development.
Some of the data presented here were obtained at the Nishi-Harima Astronomical Observatory of the University of Hyogo through the framework of the Optical and Infrared Synergetic Telescopes for Education and Research (OISTER). 
This work and operations of OISTER were supported by the Optical and Near-infrared Astronomy Inter-University Cooperation Program and the Grants-in-Aid of the Ministry of Education. 
Some of the solar data presented here were obtained at the Hida Observatory of Kyoto University, which is partly supported by Project for Solar-Terrestrial Environment Prediction (PSTEP).
We are grateful to the staff of Hida Observatory for the instrument development and daily observations.
We would like to thank A. Asai and K. Otsuji for useful discussions on solar mass ejections.
Funding for the TESS mission is provided by NASA's Science Mission directorate.
We would like to thank J.J. Drake and S. Yashiro for kindly providing the data of CMEs in their works.
We would like to thank A.F. Kowalski for carefully checking the English and contents.
We also would like to thank Enago (www.enago.jp) for the English review.
K.N. is supported by the JSPS Overseas Challenge Program for Young Researchers. 
Y.N. was supported by JSPS Overseas Research Fellowship Program. 
We acknowledge the International Space Science Institute and the supported International Team 464: The Role Of Solar And Stellar Energetic Particles On (Exo)Planetary Habitability (ETERNAL).
We thank the anonymous reviewers for their constructive comments, which helped us to improve the manuscript.
This research is supported by JSPS KAKENHI grant numbers 18J20048, 21J00316 (K.N.), 17K05400, 20K04032, 20H05643 (H.M.), 21J00106 (Y.N.), 20K14521(K.I.), 21H01131 (K.S., H.M., S.H., K.I. and D.N.) and 15H05814 (K.I.)

%\item[Online Content] Methods, along with any additional Extended Data display items and Source Data, are available in the online version of the paper; references unique to these sections appear only in the online paper.

\item[Author contributions] K.N. led the campaign observations, conducted the data analyses, and wrote the draft of the manuscript; 
H.M. and S.H. partly contributed to the data analyses;
K.N., H.M., S.H., S.O., K.I., and D.N. contributed to the observations at the Seimei telescope; 
J.T., M.T., T.O., T.S., N.K., and M.T. contributed to the observations at the Nayuta telescope;
K.I. and T.T.I. are responsible for instruments and observations at the SMART telescope;
H.M., K.L.M., F.O., M.N., R.A., M.O., and M.S. contributed to the ground-based follow-up photometric observations;
K.N., H.M., Y.N., D.S., K.I., and K.S. contributed to the data interpretation.
 
\item[Competing Interests] The authors declare that they have no competing financial interests.

 \item[Correspondence] Correspondence and requests for materials
should be addressed to Kosuke Namekata (email: namekata@kusastro.kyoto-u.ac.jp ).

\end{addendum}

\clearpage

\setcounter{figure}{0}
\renewcommand{\figurename}{\scriptsize \bf\sffamily\noindent{Extended Data Fig.}}

\begin{figure}
\begin{center}
\includegraphics[width=10cm]{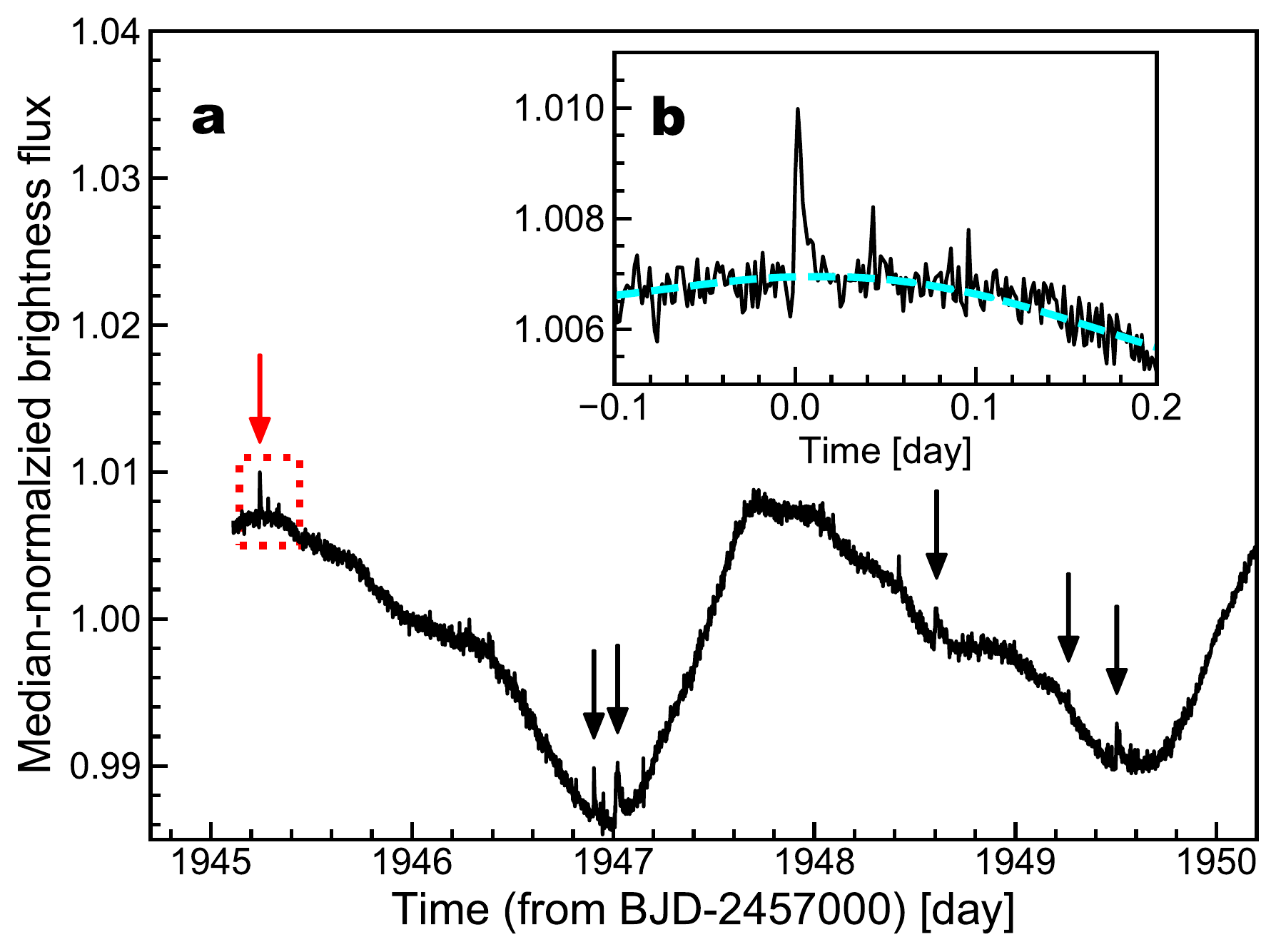}
\end{center}
\caption{
Global TESS light curve for EK Dra. \sf \textbf{(a)} Light curve of a superflare on EK Dra observed by TESS from BJD 2457000. The gap before about BJD 2458945 days corresponds to a gap in the data downlink with Earth during the spacecraft's perigee. The arrows indicate stellar flares that occurred during this observational period, but the other flares in this figure were not observed by ground-based spectroscopic observations. The red arrow is the superflare shown in Fig. \ref{Figure1}. \textbf{(b)} Enlarged light curve indicated with the red dotted box in panel \textbf{(a)}. The cyan dashed line is the global trend of the light curve caused by the stellar rotation with large starspots.
}
\label{FigS1}
\end{figure}

\begin{figure}
\begin{center}
\includegraphics[width=15cm]{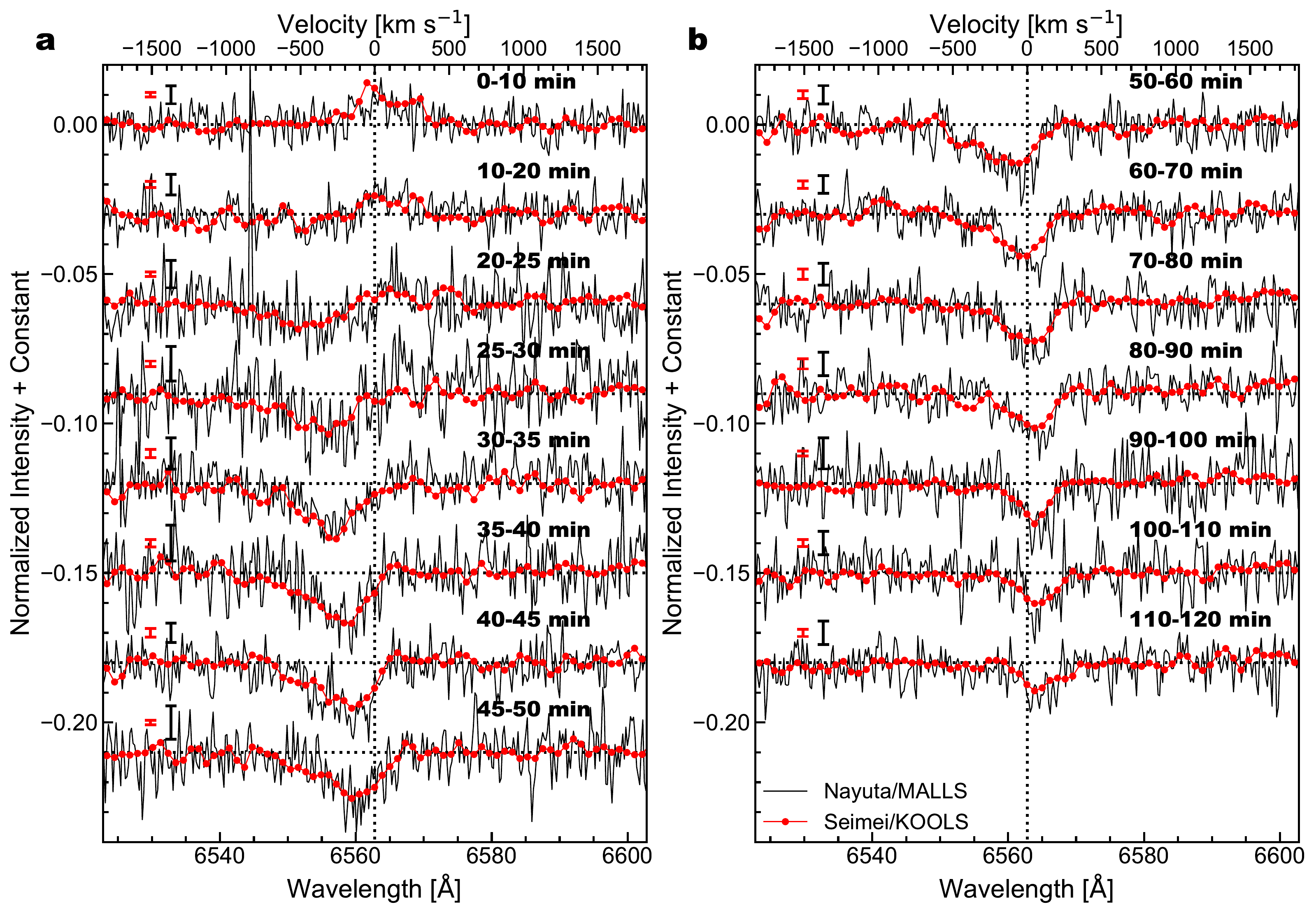}
\end{center}
\caption{
Pre-flare-subtracted H$\alpha$ spectra during and after the superflare on EK Dra with higher time cadence than panel \textbf{(e)} in Fig. \ref{Figure1}. \sf \textbf{(a, b)} The red and black lines are the data observed by the Seimei Telescope and the Nayuta Telescope. The spectra are binned in time, and the integration periods correspond to the horizontal axes of panel \textbf{(a-d)} in Fig. \ref{Figure1}.  The intensities are normalized by the stellar continuum level. The vertical dotted line indicates the H$\alpha$ line center, and the horizontal dotted lines indicate the zero levels for each spectrum. 
The 1-$\sigma$ values for the line center are indicated with red (Seimei Telescope) and black (Nayuta Telescope) error bars for each time bin.
The 1-$\sigma$ values are basically calculated by the scattering in line wing (6522.8 - 6532.8 {\AA} and 6592.8 - 6602.8 {\AA}).
 }
\label{FigS7}
\end{figure}

\begin{figure}
\begin{center}
\includegraphics[width=12cm]{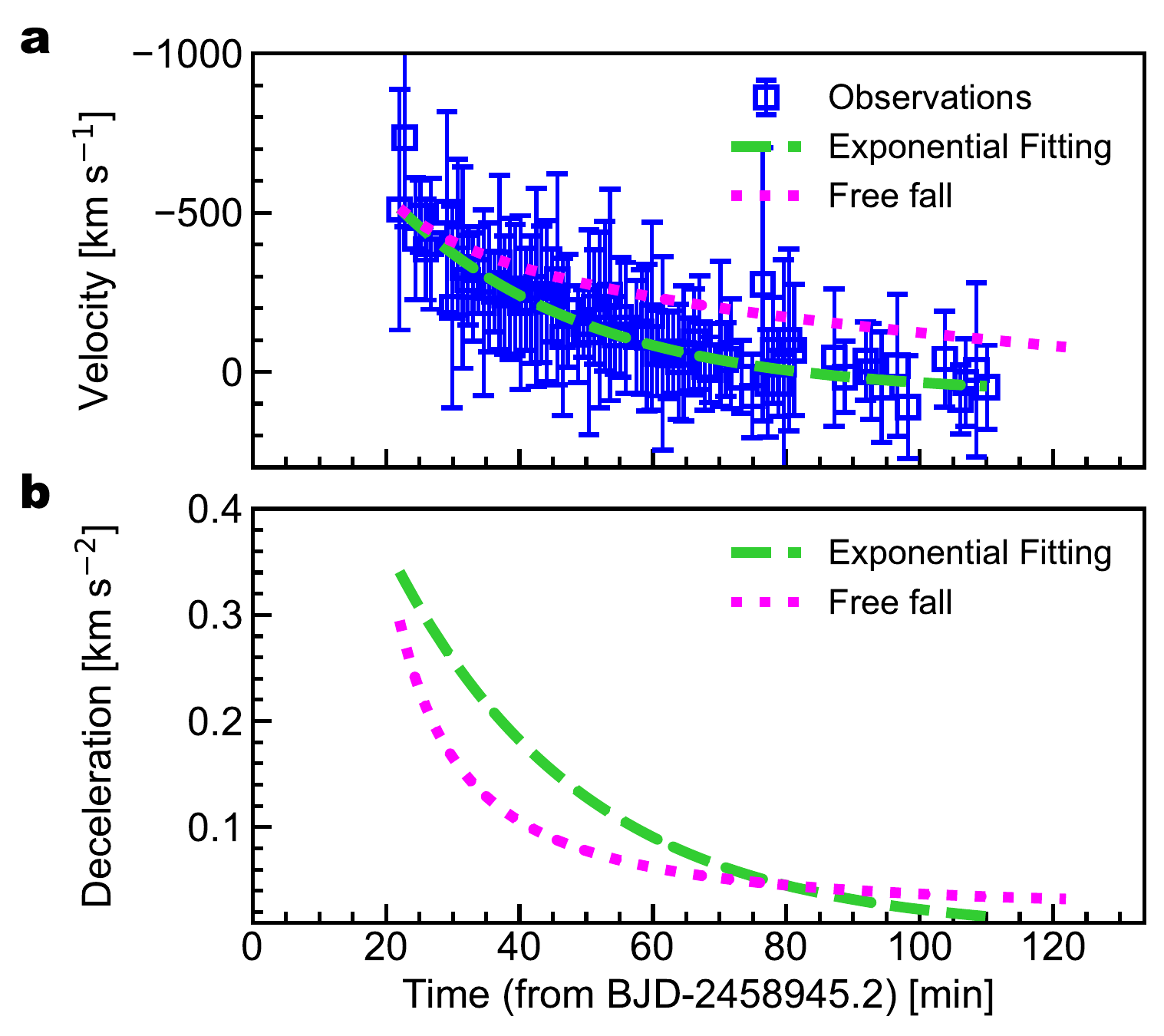}
\end{center}
\caption{
Temporal evolution of the velocity and deceleration for the H$\alpha$-line absorption features. \sf \textbf{(a)} The blue points indicate the velocity of the H$\alpha$-line absorption features seen after a superflare. The spectra observed by Seimei Telescope were used considering the high S/N and the absorption features are obtained by fitting them with a normal-distribution function. The error bars indicate the standard deviation of the fitted normal distribution. The green dashed line indicates the exponential function which fits the blue symbols, and the magenta dotted line indicates the velocity evolution of the free fall. \textbf{(b)} The temporal evolution of deceleration rates is derived from the velocity changes for observation and free-fall model in panel \textbf{(a)}.
}
\label{FigS10}
\end{figure}

\begin{figure}
\begin{center}
\includegraphics[width=11cm]{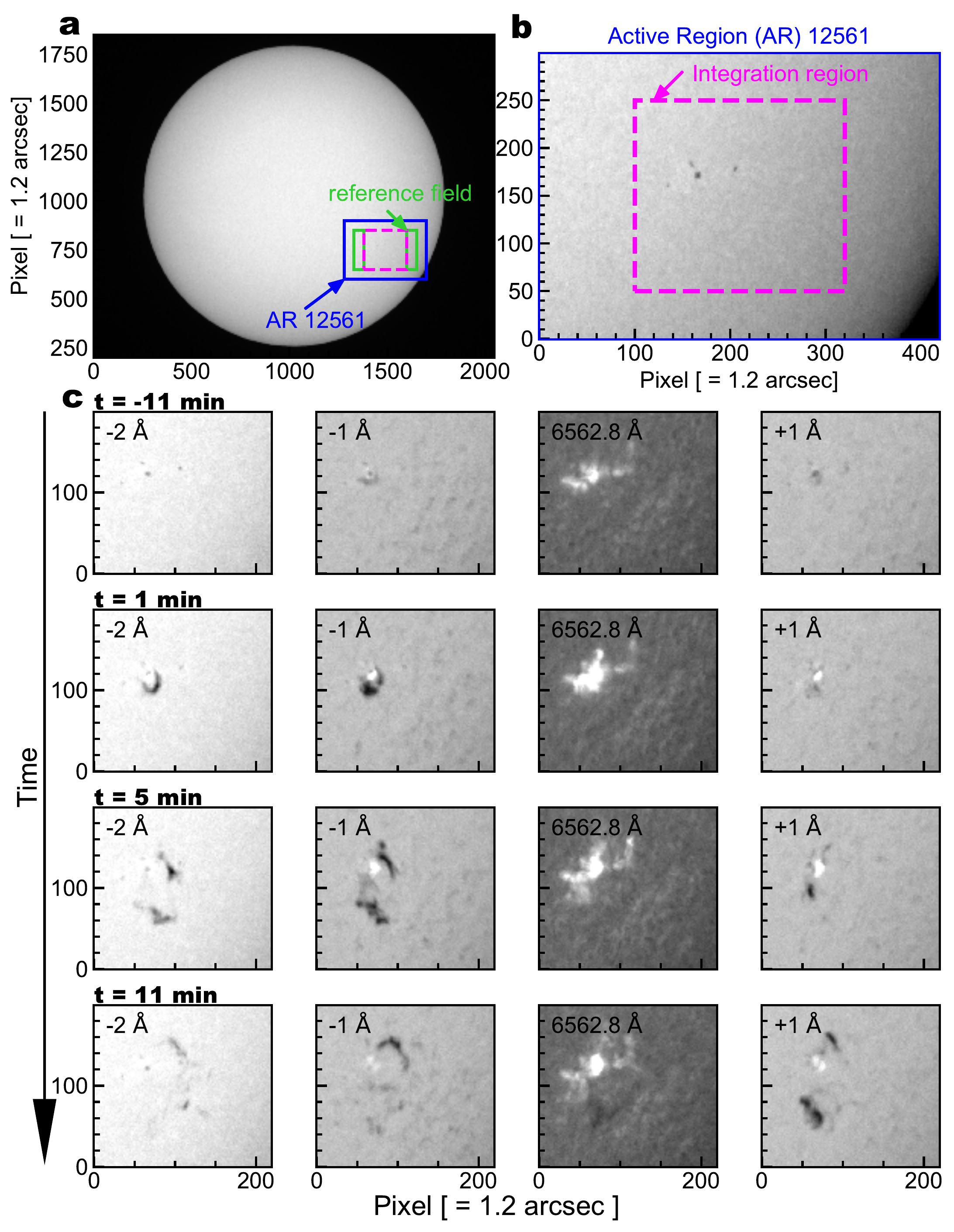}
\end{center}
\caption{
A solar flare on  7 July, 2016, observed by SMART telescope/SDDI at Hida observatory. 
\sf \textbf{(a)} A full disk image of the Sun at the H$\alpha$ line wing (6554.8 {\AA}).
The horizontal and vertical axes indicate the x-y axes in the unit of the image pixels whose size is about 1.2 arcsec.
The green region is a quiet region used as a reference to make the H$\alpha$ spectra. \textbf{(b)} the blue region is the enlarged panel of the active region 12561. 
The magenta is the region where the solar flare and filament eruption happened. 
\textbf{(c)} The temporal evolution of solar images in the magenta region at a wavelength of 6560.8 ($-$91 km s$^{-1}$), 6561.8 ($-$46 km s$^{-1}$), 6562.8 (0 km s$^{-1}$), 6563.8 {\AA} (+46 km s$^{-1}$). 
The emission and absorption features are indicated with white and black, respectively. 
The movie is available in Extended Data Movie 1.
}
\label{FigS8}
\end{figure}

\clearpage

%\renewcommand{\tablename}{\noindent{Extended Data Table }}

%The blue symbols in Extended Data Fig. 3a are fitted with the exponential function indicated with the green line, and it is converted to the deceleration rates plotted in Extended Data Fig. 3b. 

\end{document}